\documentclass[12pt]{iopart}
\usepackage{graphicx}% Include figure files
\begin{document}

\title{The knee in the cosmic ray energy spectrum: a pulsar/supernova origin?}

\author{T. Wibig}
\address{Physics Dept., University of \L \'{o}d\'{z};
The Andrzej So\l tan Institute for Nuclear Studies,
90-950 \L \'{o}d\'{z} 1, Box 447, Poland}

\author{A. W. Wolfendale}
\address{Department of Physics, University of Durham, Durham, DH1 3LE, U.K.}

\begin{abstract}
The origin of the prominent 'knee' in the cosmic ray energy spectrum at an energy of several PeV is still uncertain. A recent mechanism has shown promise, however; this involves particles from a very young pulsar interacting with the radiation field from a very young supernova remnant. The ensuing nuclear reaction of the particles with the photons by way of $e^=e^-$ production then causes the characteristic knee.

In an earlier paper we argued that the mechanism would imply only one source of a very rare type - if it were to explain the spectral shape. Here we examine the mechanism in more detail and conclude that for  even a single source to work its characteristics would need to be so unusual that the mechanism would not be possible for any known type of pulsar-supernova combination.

\end{abstract}

\maketitle

\section{Introduction}
There are only two features of the cosmic ray energy spectrum which depart from smoothness: the 'knee' at a few PeV and the 'ankle' at a few EeV. The origin of both is still subject to argument. Here we examine the first feature.

The 'knee', a comparatively sudden change of slope - a steepening - has been known since the pioneering work of Kulikov and Khristianssen \cite{10}. Most workers have attributed it to the rather sudden reduction in Galactic trapping efficiency, the trapping being confinement by the Galactic magnetic field. The different atomic nuclei will have 'cut-offs' at Z-dependent energies and, when summed, the result will be a rather smooth (not sharp) change in the spectral slope; we term this the Galactic Modulation, GM, model. A minority of workers have gone along with our idea of the knee being due to a single, recent, local supernova remnant (SNR) (\cite{3} %(Erlykin and Wolfendale, 1997
and later publications) This is the single source model (SSM). Yet another, recent, model has been put forward by Hu et al. in Ref.~\cite{7}
%(2009)
(to be referred to as I) in which particles from a rapidly rotating pulsar interact with radiation from the parent SNR. We refer to this model as PSNR.

The aim of the present work is to follow up our earlier work
(\cite{4}),
%Erlykin et al. 2010),
(to be referred to as II) in which we examined the PSNR Model as originally proposed in I to see if the knee caused by the particle -SNR photon interactions was a universal property of all pulsar, SNR pairs. We found that it was not but that there was hope for a single rare pair having the required properties. This would be identified as the single source of the SSM although the mechanism would differ from that in the original SSM which involved SNR alone. Here we examine the needed properties of the pulsar, SNR pair in more detail.

\section{The basis of the PSNR model.}
\subsection{The characteristics of the interactions.}
As remarked already, the philosophy of the source process is a composite pulsar-SNR system with the pulsar-accelerated particles interacting with the photon field of the SNR. Under certain circumstances a sharp knee will result due to particle, photon $\rightarrow$ e$^+$e$^-$ interactions. We start by examining the general properties of the interaction in terms of the mean energy of the photons.

\begin{figure}[th]
\begin{center}
\centerline{\includegraphics[width=7.9cm]{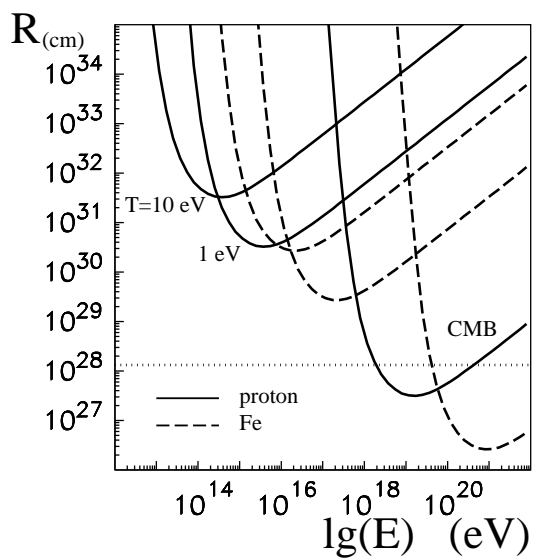}}
\end{center}
\caption{Absorption length vs energy for protons and iron nuclei interacting with photon fields of mean 'temperature' as shown. The energy density of all but the CMB is 1 eV cm$^{-3}$}
\label{fig1}
\end{figure}

An important characteristic of the 'knee' in the energy spectrum is its 'sharpness', $S$. We follow
Ref.~\cite{3} %Erlykin and Wolfendale (1997)
and define the sharpness as
\begin{equation}
S\ =\ -\ {\partial^2 \:(\log IE^3) \over \partial\: ( \log E)^2}.
\end{equation}
Figure \ref{fig1} sets the scene for the attenuation length, $R$, of protons and iron interacting with black body photons. The energy density of the photons is 1~eV~cm$^{-3}$ for all but the Cosmic Microwave Background (CMB), where it is the usual 0.24~eV~cm$^{-3}$. The minimum range is inversely proportional to the mean photon energy, $T$. (1 eV corresponds to $1.16 \times 10^4$~K, ie a mean wavelength of 12,400 \AA). We note that the middle of the visible range, yellow light, has wavelength $\sim 5000$~$\AA$ ie a mean energy of 2.48~eV.

The effect of the photon field on the energy spectrum of the emerging particles depends on the energy density of the photons at the time of their passage through the field, for a low energy density there is no effect and for a high energy density the interactions are catastrophic, with no particles emerging. As an illustration, with the object of finding $S$-values in the required range we choose a combination of energy density and 'trapping time' such as to give the minimum $R$ in Fig.~\ref{fig1} for $T = 1$~eV ($R = 3 \times 10^{30}$~cm). This must be equal to $c \int E(t)~t ~ dt$, where $E(t)$ is the energy density at time $t$. The limits, $t_0$ and
$t_1$, are discussed in the next section. So as to maximise the PSNR 'effect' we choose the smallest value of $t_0$ and the largest $t_1$ permissible.

\subsection{The limiting time parameters for the interactions and the two models.}
The lower limit to the interaction time, $t_0$, is set by energy losses by the pulsar-accelerated particles on the environmental gas density. This latter is enhanced very considerably by the ejecta from the SN; it is this material, too, that obscures the optical radiation for a period of days. For a typical massive progenitor of a Type II SN (and a 'magnetar') the mass ejected is of order 5~$M_\odot$ \cite{8}
%(Kasen and Bildsten, 2010)
The expansion velocity at short times after the expansion can be found by reference to model calculations only; a commonly quoted value is $\sim c/10$ Some support for a value of this order comes from [5A], who measured c/30 some 30 years after the expansion of a shell remnant (43.31, +592) in Messier 82.

If the density of the ejecta were constant, an attenuation length for protons of 120~g cm$^{-2}$ would be achieved in $\sim 1.6 \times 10^6$~s, neglecting the 1SM swept up. If concentrated in a thin shell the time needed would be $\sim 10^6s$. In fact, there will be considerable clumpiness, reducing this time by at least an order of magnitude; indeed, observations of (much older) SNR show very variable densities. Here, as a rather extreme limit (a factor 30 smaller) we adopt $t_0 = 3 \times 10^4 $~s. A measure of support comes from the fact that, occasionally amongst the very many SNR light curves one encounters `light break through' at such early times.

Turning to the maximum time, $t_1$, its' derivation is easier. Inspection of Figure~\ref{fig1} shows that for a photon 'temperature' of 1eV and an energy density of 1~eV~cm$^{-3}$, interactions will cease when R $\sim 10^{30}$cm. This can be converted to an appropriate radius for the environment of a SNR of known total photon energy and its time dependence.The highest energy output achievable is $\sim 10^{43}$~erg~s~$^{-1}$ \cite{14}
%%Nakar and Sari, 2010)
yielding in 10$^6$~s : $10^{49}$~erg.

We now come to an important problem: the extent to which the pulsar-accelerated particles are trapped in the remnant before escaping.
Alternatives are adopted here:
\begin{itemize}
\item[(i)] the particles are trapped by the tangled magnetic fields and diffuse out with the remnant (Model 1) and
\item[(ii)] the particles are not trapped but stream out from the pulsar with the speed of light (Model 2)
\end{itemize}

Predictions will be made for each case and then common problems will be addressed.

\section{Results for Model 1.}
\subsection{Model 1 - particle diffusion.}
SNR commonly expand at a velocity of $\sim c/10$ as already remarked and an expansion time of $10^6$~s yields a radius of $3 \times 10^{15}$~cm. Assuming that the CR are trapped in the remnant for this time the required total distance traveled for the photon energy achieved corresponds to $R \simeq 10^{30}$~cm for 1~eV~cm$^{-3}$. The energy density for our remnant follows as $\simeq 5\times 10^{13} $~eV~cm$^{-3}$.

To summarise: our (rather extreme) limits are $t_0 = 3 \times 10^4$~s and $t_1 = 10^6$~s; the corresponding SNR radii are 10$^{14}$~cm and $3 \times 10^{15}$~cm.

\subsection{Model 1 - The derived spectrum after the interactions.}
\begin{figure}[th]
\begin{center}
\centerline{
\includegraphics[width=7.9cm]{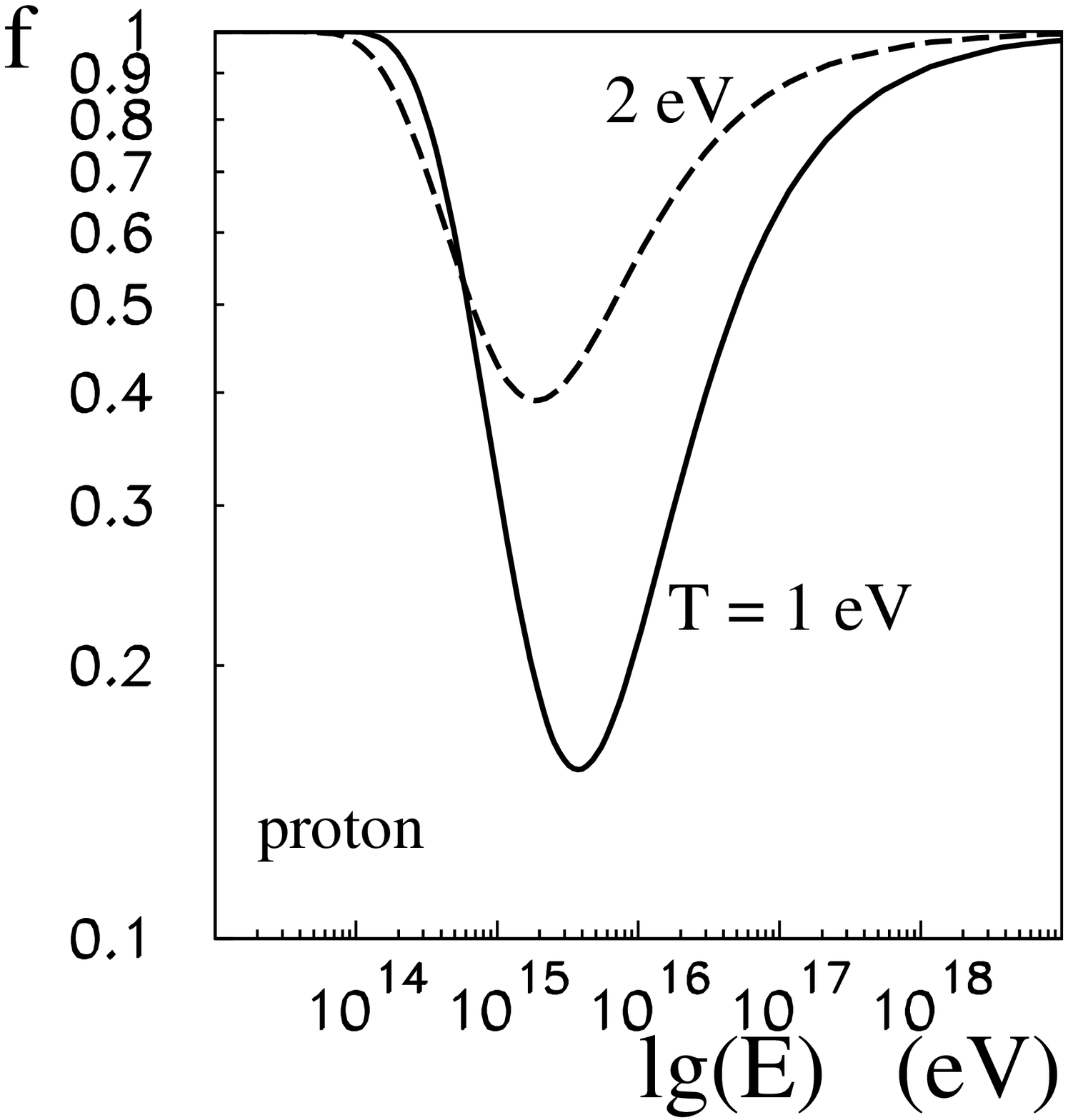}
\includegraphics[width=7.9cm]{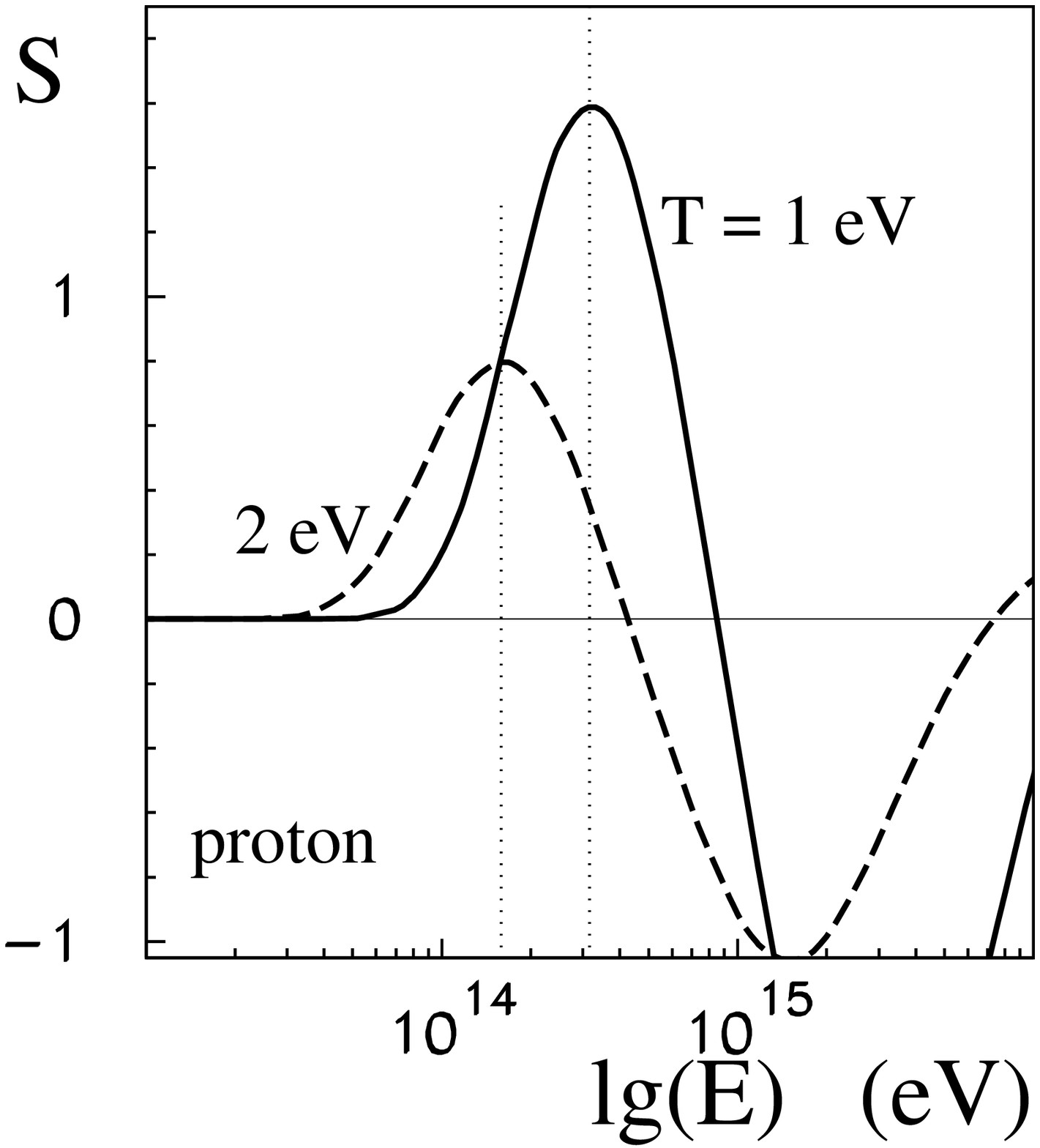}}
\end{center}
\caption{(a) The reduction factor, to be applied to the proton spectrum due to
interactions with the photon field chosen (see text) for two 'temperatures'.
(b) Sharpness values derived from the above.
}
\label{fig2}
\end{figure}

Calculations have been made using the above parameters, for protons in the first instance. The results are shown in Figure~\ref{fig2}, for both the spectral reduction factors, $f$, and sharpness values, $S$. In order to study the sensitivity to the mean photon energy, the calculations are made for both $T = 1$~eV and 2~eV.

It should be noted that the $S$-values are upper limits in that we ignore the particles which are accelerated after the interactions with the SNR-photons have effectively ceased. Such particles will dilute the knee and reduce $S$.

\begin{figure}[bh]
\begin{center}
\centerline{
\includegraphics[width=7.9cm]{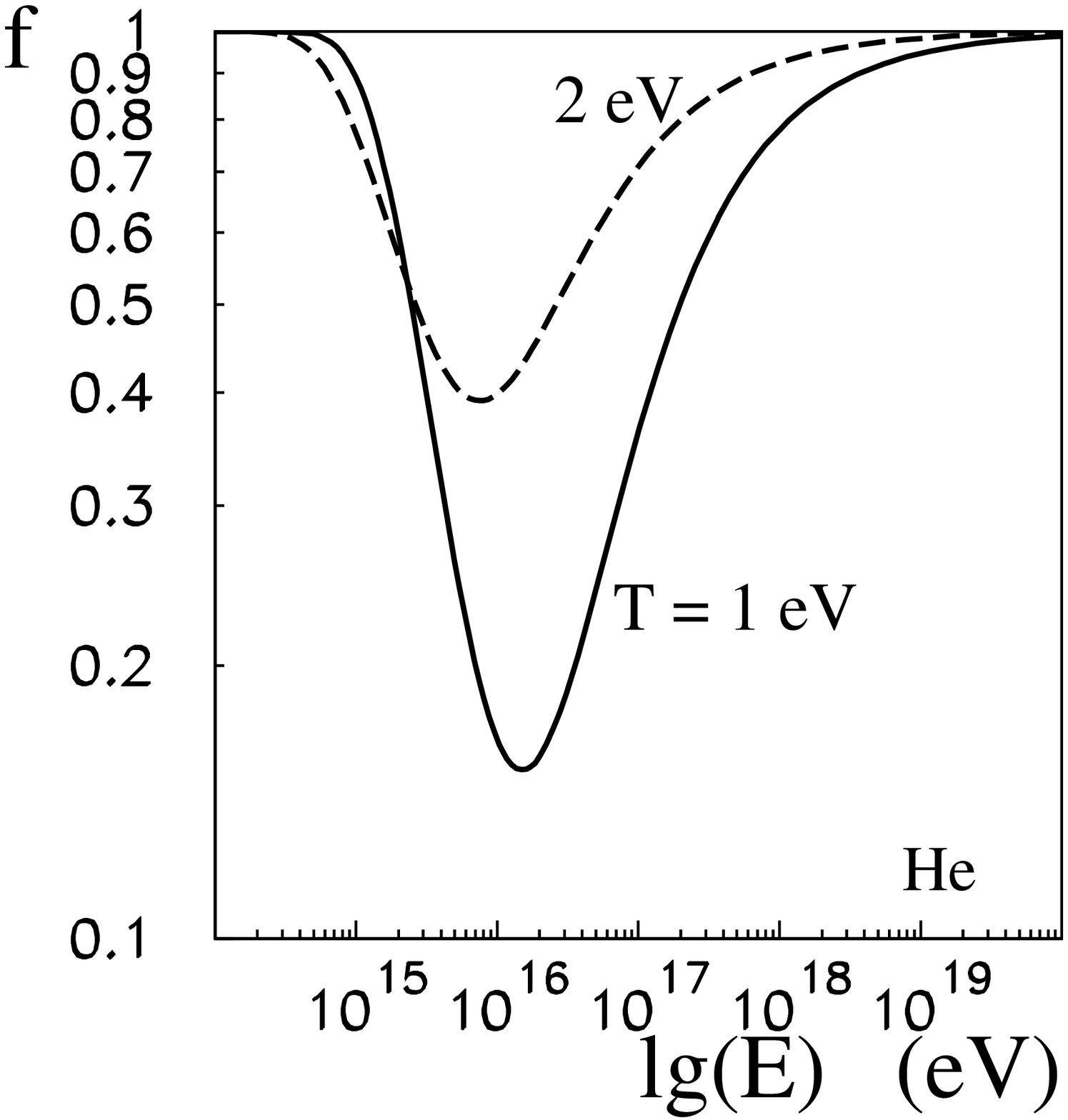}
\includegraphics[width=7.9cm]{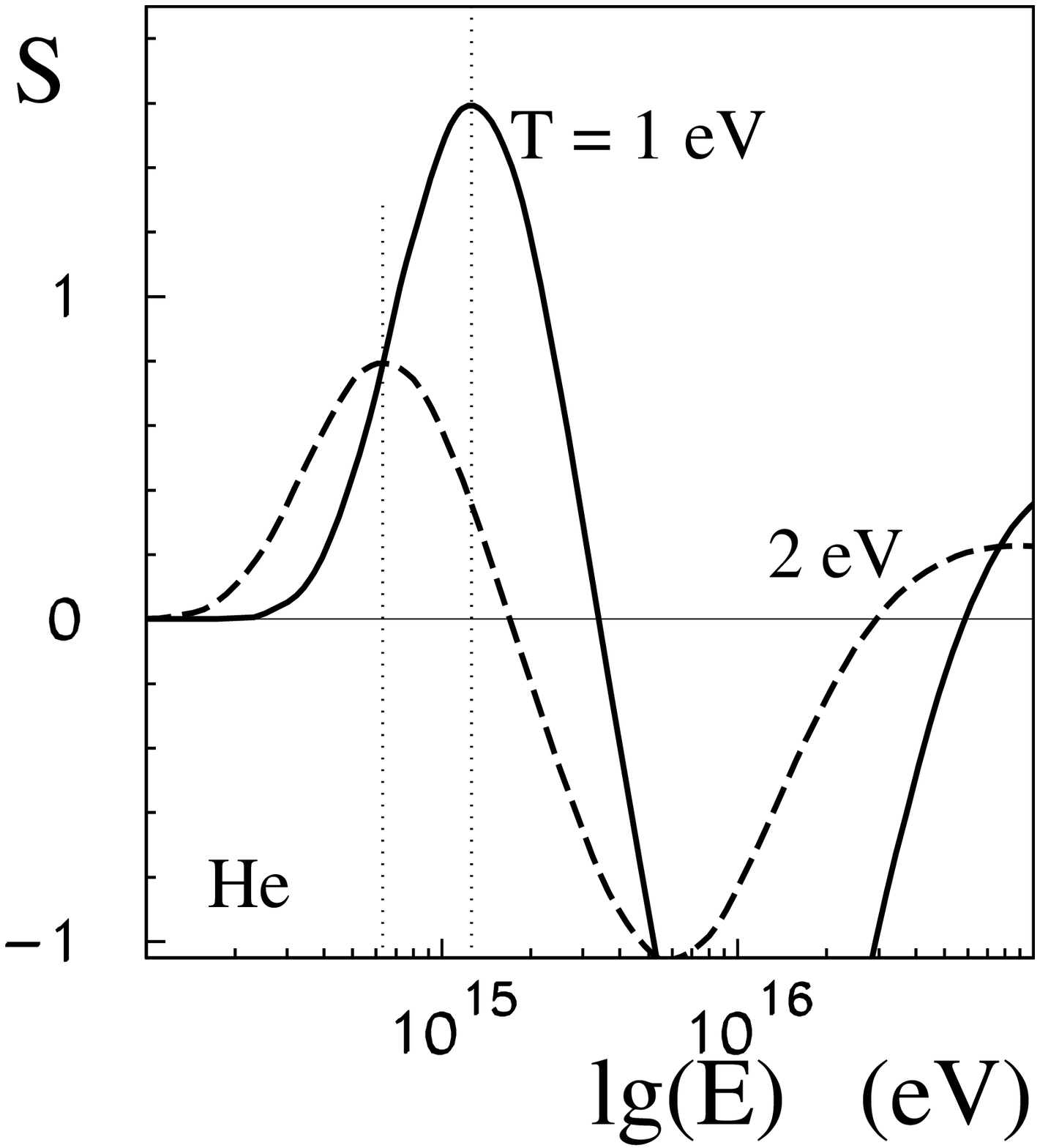}}
\end{center}
\caption{(a),(b) As for Figure 2 but for Helium nuclei.}
\label{fig3}
\end{figure}

Figure~\ref{fig3} gives the corresponding plots for Helium; it will be seen that the $S$-values are rather similar - the main difference being, as expected, in the energy at which $S$ is a maximum.

\subsection{Model 1 - The ensuing knee position}
Inspection of Figures~\ref{fig2} and \ref{fig3} shows that the energies for $S_{\rm max}$ are, for protons:
$E^*$(p): 0.31 PeV (1~eV) and 0.15~PeV (2~eV)
and for helium nuclei:
$E^*$(He): 1.2 PeV (1~eV) and 0.62~PeV (2~eV)

The ratio of the energies for $S_{\rm max}$ for helium and protons are seen to be $\sim 4$, as would be expected since it is energy per nucleus that is important
in determining the threshold energy for photon, particle
interactions.

Comparison can now be made with experiment. The energy of the knee in the CR energy spectrum is model-dependant and values range from about 3 -- 8 PeV. There is evidence that Helium nuclei predominate at the knee
\cite{1},
%(Apel et al., 2009).
\cite{5}).
%Erlykin and Wolfendale, 2010).
If so, there is clearly a factor 10 difference in the $S_{\rm max}$-energies, between observed and expected, for $T$ = 1~eV. A value nearer $T$ = 0.1~eV (wavelength $\sim  120,000$ \AA: for infra-red) is required. This is a serious drawback for Model I.

%Turning to the ratio of the knee energies, the model gives $E^*$(He)$/E^*$(p)$=4$,
%in comparison with 2 for rigidity-dependant (eg SNR shock wave) acceleration.

%Comparison can be made with observation. Only the KASCADE EAS results have been 
%analysed in such a way as to determine the energy spectra of individual
% nuclei (protons, Helium etc.) with the result that 
%$E^*$(p) =
%(2.5$\pm$1.0)~PeV, and $E^*$(He) = (6.0$\pm$1.5)~PeV. The 
%ratio $E^*$(He)/$E^*$(p) is thus 2.4$\pm$1.1, a value nearer that for the 
%SSM than the model under consideration.

%A bigger problem is the actual values of $E^*$, as distinct from their ratios 
%(He to p), for the value of $T$ = 1~eV chosen; $E^*$(p) is a factor 8 too small, 
%and for the alternative of $T$ = 2~eV the factor is 17.

\subsection{Model 1 - A variant to fit the CR spectral knee.}

\begin{figure}[th]
\begin{center}
\centerline{
\includegraphics[width=7.9cm]{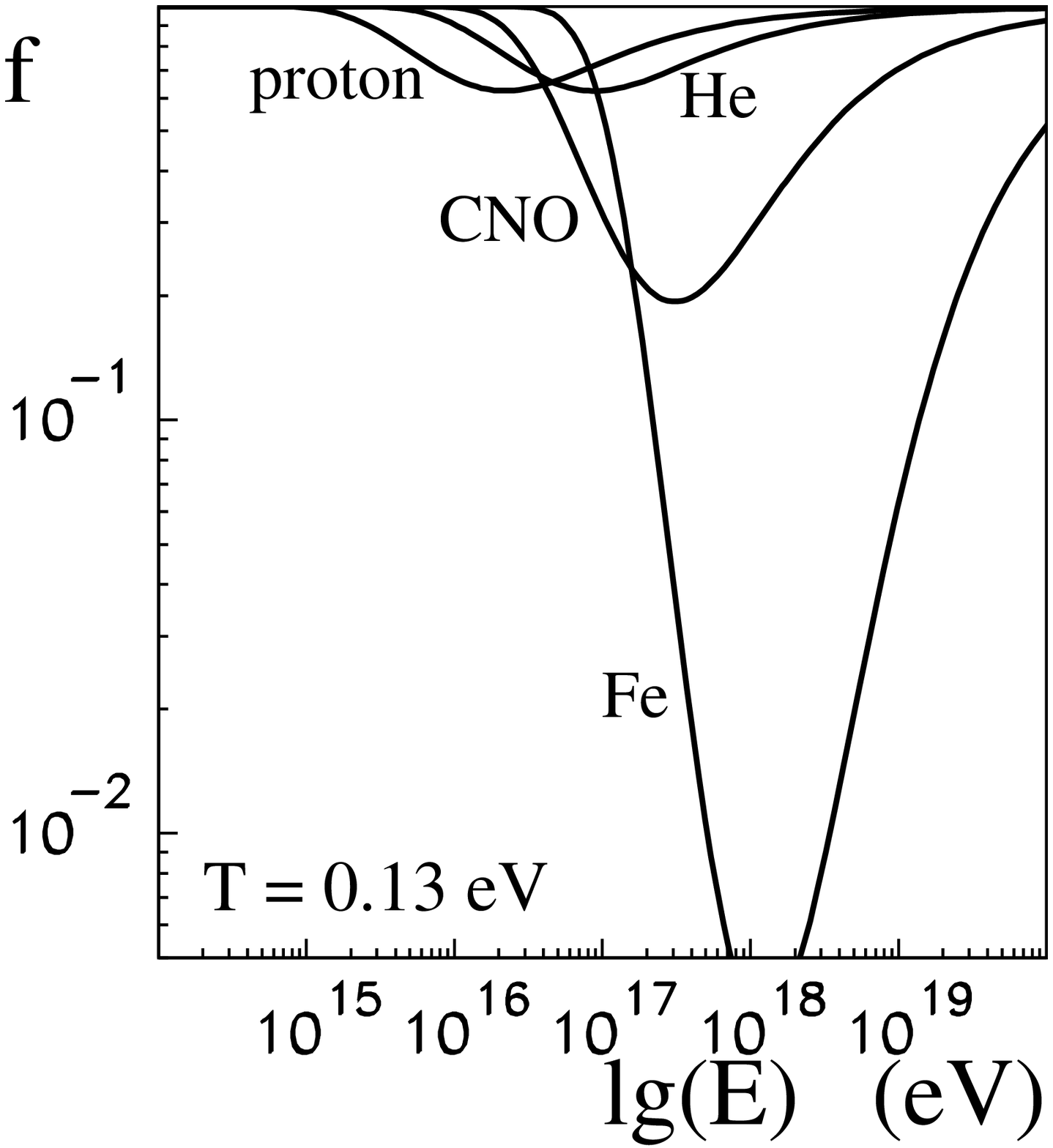}
\includegraphics[width=7.9cm]{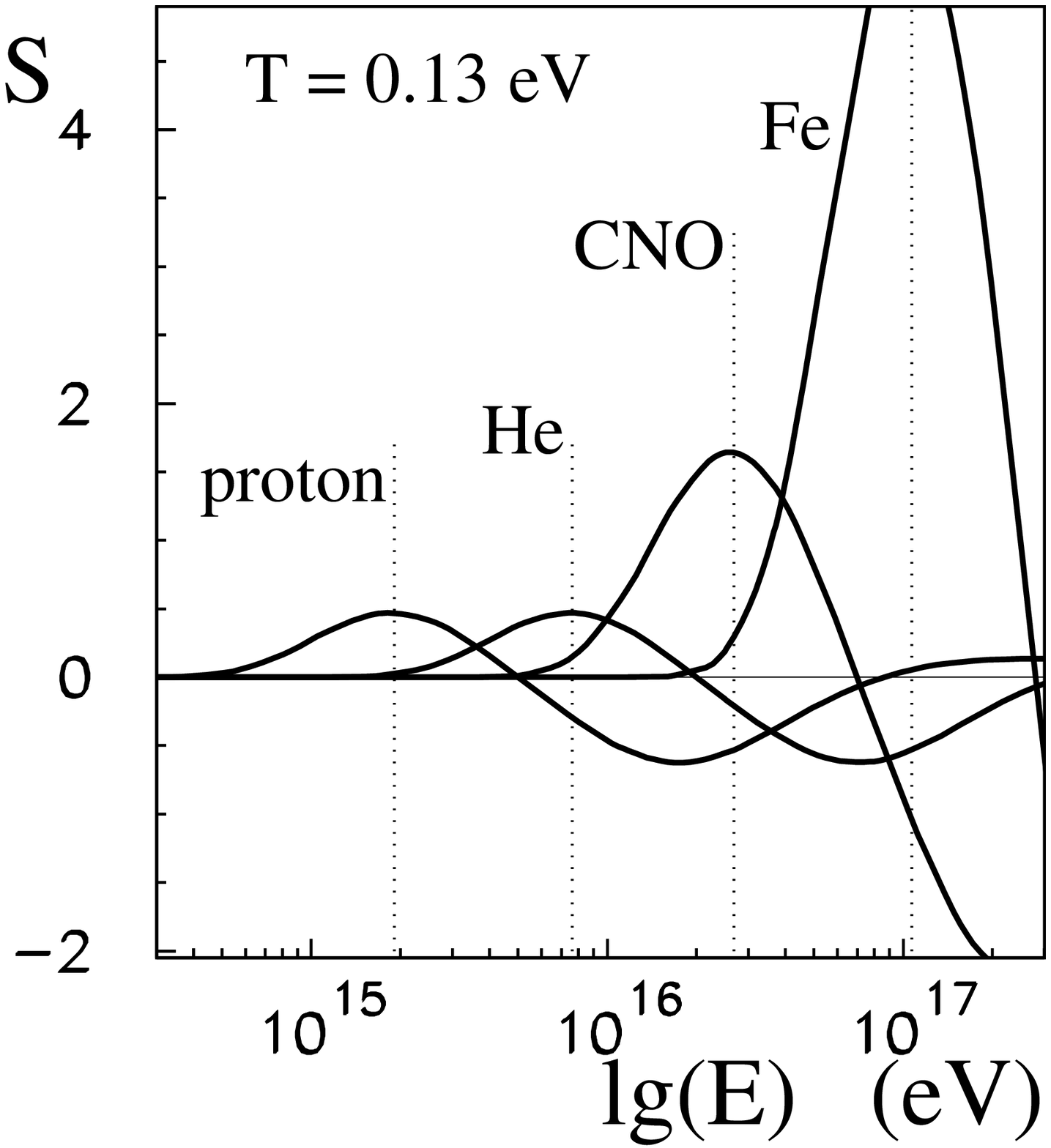}}
\end{center}
\caption{(a) $f$-values for the nuclei shown as a function of particle energy for $T = 0.13$ eV, the temperature needed to give knee positions for protons and He near those found experimentally.
(b) sharpness values derived from the above.}
\label{fig4}
\end{figure}

Figure~\ref{fig4} shows the situation where $T$ is reduced to 0.13 eV so that the proton knee (peak) is now at $E^*$(p) $\simeq$ 2 PeV. The Helium knee (peak) is at $E^*$(He) $\sim$ 7.8 PeV. Both can be regarded as, perhaps, just acceptable.

Two aspects need consideration, now. The first concerns the position of the iron knee (peak), at 100 PeV. This is close to the 70 PeV where recent work \cite{5} gives rather strong evidence for a peak from an analysis of the results from the EAS arrays. This is a good feature of the model. The second relates to the need for $T\sim$ 0.13~eV; with the conversion 1 eV corresponds to $1.16 \times 10^4$~K, the effective temperature is in the region of 1,500 K, ie in the infrared, as already remarked. Although there is, inevitably, an infrared flux, from visible radiation absorbed by the SNR material, its intensity is too low to give sufficient energy density for our purpose.

\subsection{Model 1 - The effect of a non-uniform SNR temperature.}
\begin{figure}[th]
\begin{center}
\centerline{
\includegraphics[width=7.9cm]{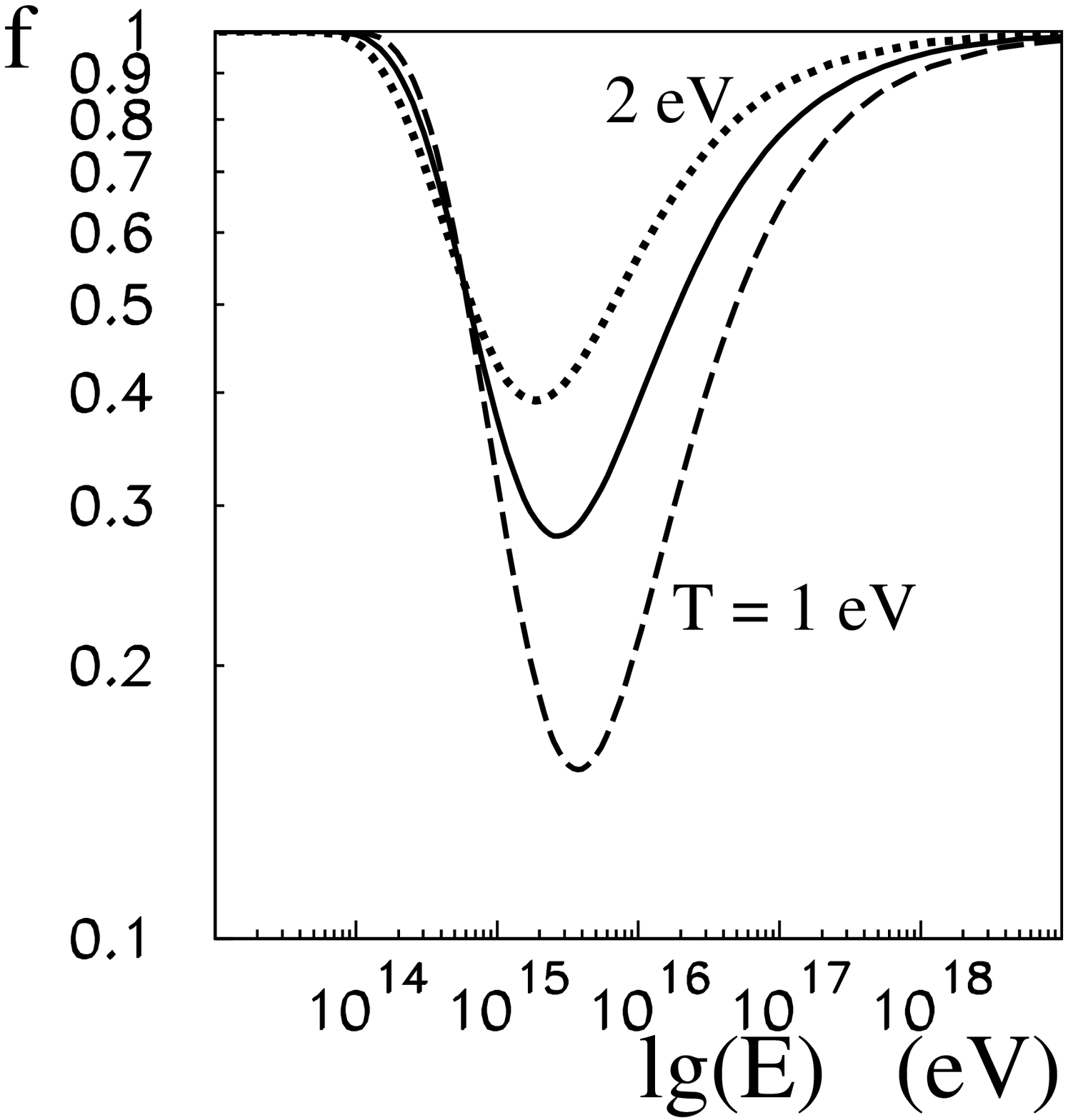}
\includegraphics[width=7.9cm]{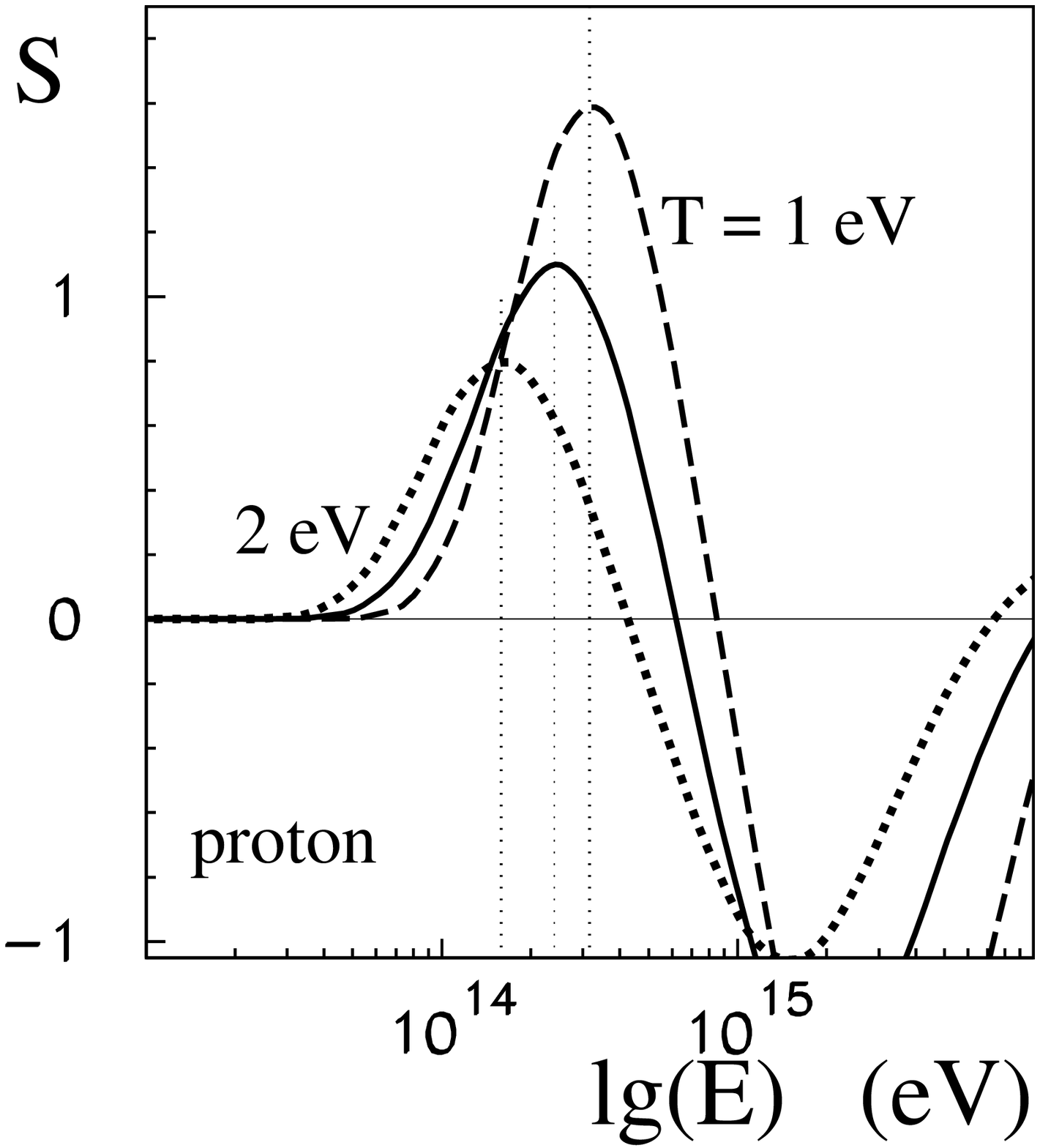}}
\end{center}
\caption{(a) $f$-values for a model in which the $T$-value fell smoothly from 2 eV to 1 eV during the interaction period (full line), in comparison with constant values of 2 eV and 1 eV.
(b)Sharpness values derived from the above.}
\label{fig5}
\end{figure}

Inevitably, the temperature of the remnant will fall with time and a reduction of $S$ will occur. Figure~\ref{fig5} shows the situation for the likely case of a 'conventional SNR' (eg \cite{9}) where the temperature falls by a factor 2 over the operative period of particle-photon interactions. Interestingly, although finite, the reduction in $S$ is not large. However, for more dramatic SNR, such as are required here, the change in temperature can be much bigger. In \cite{14} temperatures versus time are given for the following progenitors: red supergiants, blue supergiants and Wolf-Rayet stars. The ratios of $T$($3\times 10^4$~s) to $T$($10^6$~s) are, respectively, (for typical stars within these types): 7.1, 8.4 and 8.4. The exponents of the power laws are -0.56, -0.61 and -0.61. It is true that the luminosity falls with time, too, but with much smaller exponents: -0.17, -0.35 and -0.35. In consequence, the reduction in $S$ in practice for SNR from such progenitors will be much smaller than shown in figure~\ref{fig5}.

For young magnetars, potential SN for the mechanism considered here, the light curves are a rather strong function of the magnetic field of the magnetar and often continue for hundreds of days \cite{5}. Predictions are not available for the short period from SN birth (0.3 to 10 days) needed here so that conclusions cannot be drawn.

In summary, falling temperatures appear not to be a problem for the model.

\section{Results for Model 2}
%As remarked earlier, in this model the particles stream out from the pulsar with
% the speed of light. The observed SNR light curves, with their delayed maximum
% output indicate that the photons are themselves 'trapped' in the expanding SNR 
%shell, however. To quantify this we adopt model luminosity functions (L vs time) 
%from the exploding SN from Ref.%the work of Nakar and Sari \cite{14}
%(2010)

%$$
%\begin{array}{l}
%L(t)~=~\left\{
%\begin{array}{rcl}
%10^{44}\:t_{\rm days}^{-1.33} & {\rm egr/sec} & t < 14 ~{\rm h} \\ \   \\
%3\times 10^{32}\:t_{\rm days}^{-0.17} & {\rm egr/sec} & t > 14 ~{\rm h}
%\end{array} \right .
%$$
%\\
%\
%\\
%$$
%T(t)~=~\left\{
%\begin{array}{rcl}
%10 \:t_{\rm hours}^{-0.36} & {\rm eV} & t < 14 ~{\rm h} \\ \   \\
%3  \:t_{\rm days}^{-0.56} & {\rm eV} & t > 14 ~{\rm h}
%\end{array} \right .
%\end{array}
%$$

%The functions relate to red supergiants, which are the likely precursors for the SNR 
%under consideration.

%It is assumed that the trapping has an efficiency, $\xi$, which decays exponentially 
%with a cooling time, $T_c$. The pulsar output ('source activity time' $T_a$) is a 
%crucial quantity. Calculations have been made by us for a wide range of values of 
%$T_c$ and $T_a$ with the percentages of energy absorbed as parameter and the ensuing 
%sharpness determined. Figure~\ref{fig7} shows the result for those situations which 
%give spectra of sharpness, $S > 2$: the area within the curves is the `allowed' 
%region. Clearly, for a small fraction of energy absorbed the allowed 'area' is very 
%small.

\begin{figure}[th]
\begin{center}
\centerline{
\includegraphics[width=7.9cm]{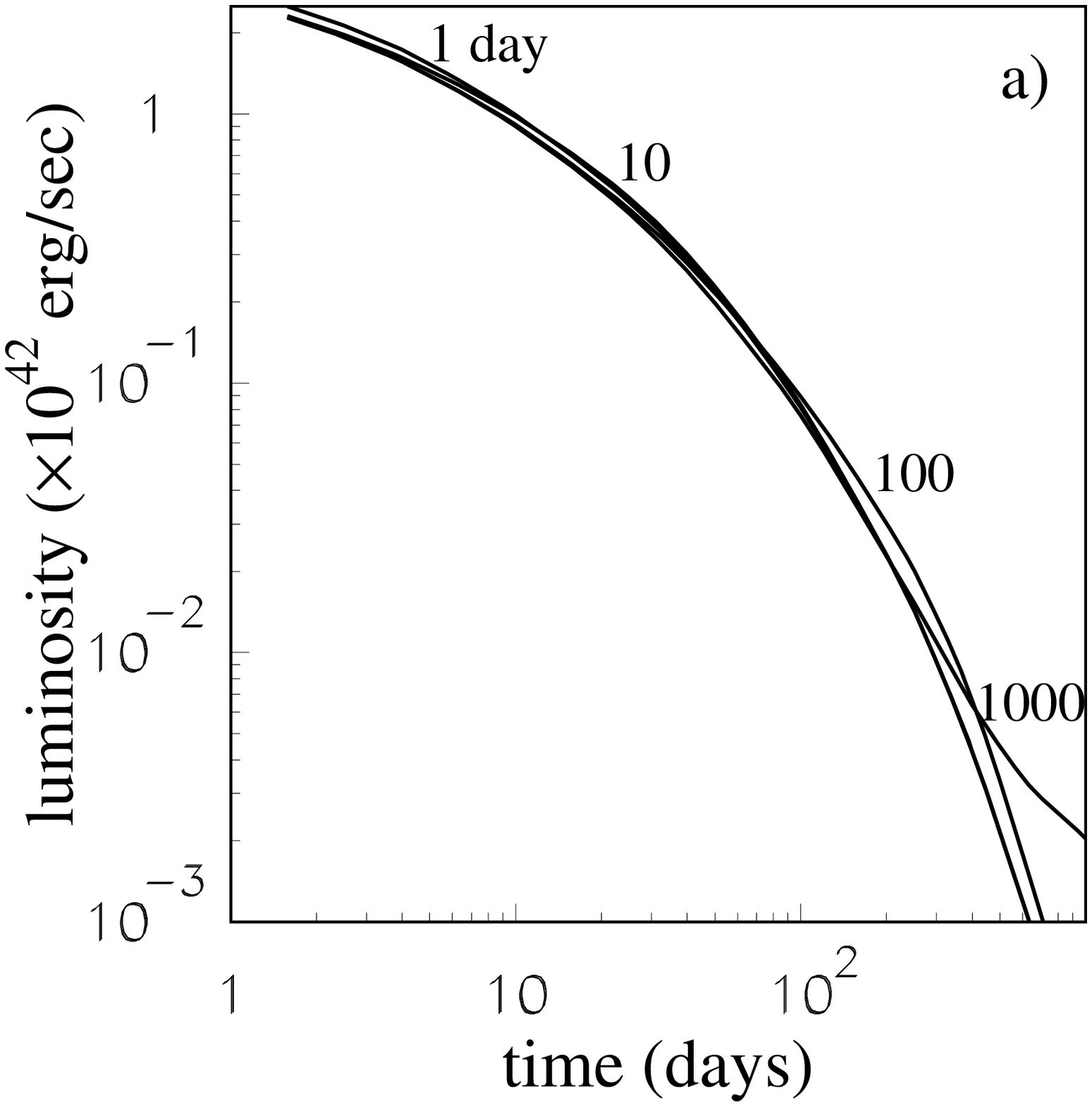}
\includegraphics[width=7.9cm]{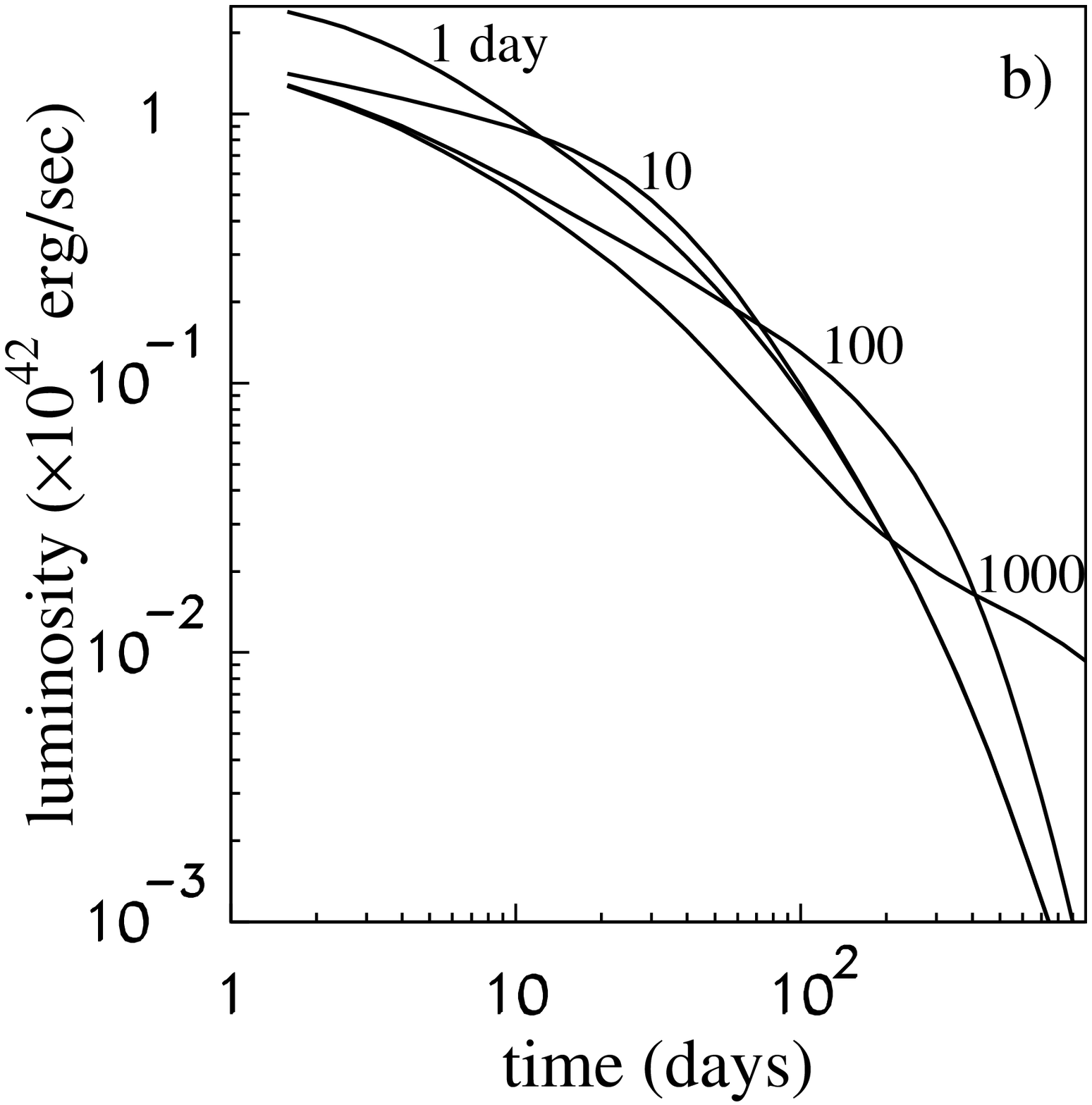}}
\end{center}
\caption{Luminisity for 10\% (a) and 50\% (b) absorption.}
\label{fig6}
\end{figure}

    Inspection of many published SNR `luminosity curves' shows a rather
wide spread but Figure~\ref{fig6}a shows the general form,for the case where the
absorption by the SNR shell material is only 10\%.
Figure~\ref{fig6}b shows the light
curve for an absorption of 50\%.
The parameter for the lines is the `cooling time',
i.e. the mean exponential decay time of the radiation.

\begin{figure}[th]
\begin{center}
\centerline{
\includegraphics[width=7.9cm]{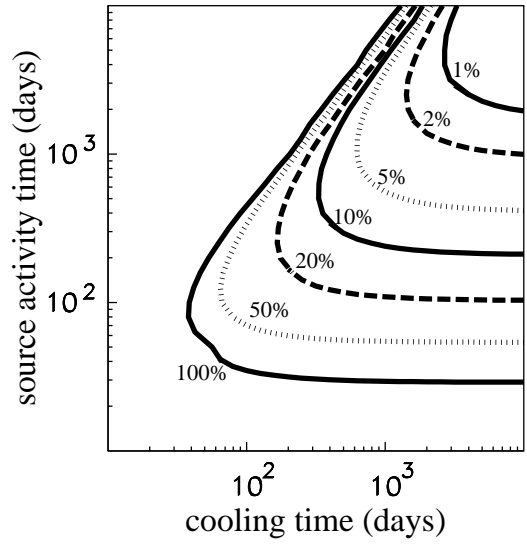}}
\end{center}
\caption{Model-2 in which the pulsar-accelerated particles travel in straight lines. The percentages refer to the fraction of the SNR radiation absorbed by the ejecta shell. 'Source activity' relates to the time for which the pulsar is accelerating particles of the needed energy. The results are for helium.}
\label{fig7}
\end{figure}

      Calculations have been made by us for a variety of values for the
absorption efficiency, the cooling time and the time over which the source is
active (the `source activity time').Demanding that the sharpness should
exceed S=2 gives the results shown in Figure~\ref{fig7}.
It will be noted that there
is a minimum cooling time of 40 days and the source activity time must be
greater than about 30$+$ days.

\section{Discussion.}
\subsection{Parameters to be discussed.}
A number of parameters need examination, principally:

\vspace{.4cm}
\hspace{1cm}
\begin{minipage}{15cm}
the masses needed for injection,\\
the total photon energy,\\
the products of the photon-nucleus interactions and, most importantly,\\
the characteristics needed for the pulsar.\\
\end{minipage}
These parameters will be considered, in turn.

\subsection{The masses needed for injection}
It is apparent that the masses of nuclei needed: protons, Helium, Iron (?) are not expected as products of pulsar acceleration. Rather, proton-rich or iron-rich neutron-star surface elements would be expected. Although not an impossible situation, the needed composition does not add confidence to the model.

\subsection{The total photon energy}
The total photon energy in the photon field adopted (which we assume effectively ceases after $10^6$~s) is $\sim 10^{49}$~erg. To this should be added the later energy - which is the energy 'seen' after the obscuration has ceased, but here we ignore this component which should not be large. The value can be compared with the total energy released in a typical SN: about
$10^{51}$ erg, of which perhaps $3\times 10^{49}$ erg appears in visible light \cite{9}.
%(Kasen et al., 2006).
Thus, the energetics are acceptable.

\subsection{The products of the nucleus-photon interactions.}
If the whole of the CR spectrum were due to the PSNR mechanism then the overall change of slope of the primary CR spectrum would correspond to a transfer of energy from CR particles to gamma rays, via the electron pairs generated in the particle-SNR photon interaction. Inspection of the CR spectrum shows that the energy density of Galactic gamma rays would be about $2\times 10^{-5}$ eV cm$^{-3}$, a value comparable with that detected in the diffuse gamma radiation, the latter usually being attributable to CR particles interacting with gas in the ISM. Such a situation would be 'dangerous' for the model but probably not catastrophic. In the more likely situation (in our view) that it is only the Single Source that produces the knee then copious gamma ray fluxes will have been produced soon after the interaction (some few 10$^5$ y ago?). However, it is likely that these gamma rays will have 'been and gone' by now.

The conclusion is that the non-observation of the gamma rays associated with the P-SNR interactions does not pose a threat to the model.

\subsection{The characteristics needed for the pulsars.}
The first characteristic is that the pulsar should be able to accelerate particles to rigidities of some 10PeV. Ref.~\cite{6}
%(2001)
gives the expression
$$R_{max} = 6.6 \times 10^{12} B_{12}\ P \ V$$
Where B$_{12}$ is the magnetic field in 10$^{12}$~Gauss, P is the Period in seconds and V is the rigidity in Volts. If we need R$_{max} = 10^{16}V$ then, for B$_{12} = 3$ (the mean, from 'model A' of \cite{11}) a period of 45~ms is required. Interestingly, this is not too far from the birth period in the same model, which has P = 23~ms.

The problem here is the likely age of pulsars of such characteristics. Using the work of
\cite{2}
% Bednarek and Bartosik (2005)
again, this is given by
$$T = \frac{P_{\rm ms}^2}{2 ~B_{12}^2} \times 10^9 = 7\times 10^{10}~{\rm s}$$

Such a time (1000~y) would be far too long to satify the requirements of Figure~\ref{fig3} (where the 'cooling time' is unlikely to be longer than $\sim$ 100~d.

Pulsars of much shorter period are required, for example, $P \sim 1$~ms and, say, $B_{12}=3$, for which $T \sim 500$~days.

'Normal' pulsars (Lorimer, Model A, \cite{11}) have a standard deviation about the mean of $\sigma(\log B) \sim 0.3$ and will have a similar standard deviation for period \cite{12}, thus they will not satisfy the bill, even to the extent of 3 standard deviations from the mean.

Magnetars \cite{13} would, at first sight, appear to satisfy the
requirements. They have extremely high magnetic fields, typically $6 \times 10^{14}$~G but, when measured, their periods are of order seconds. In \cite{14} a theory is developed for the evolution of magnetars, in which the important birth period is in the range 10-740~ms, with a mean of 161ms. However, the mean field is given by $\langle \log B \rangle \simeq 11.9.$ Clearly, the $T$-value will be, again, too long.
Another factor militating against a magnetar as being the 'local' pulsar responsible for the knee (and the electron results) is the fact that their birth rate is quoted as $\sim 10^{-3}~{\rm y}^{-1}$ (\cite{13}) furthermore none is in our vicinity; the catalogue, \cite{13}, gives the nearest as being some 3.5kpc away. In fact
the most likely nearby SNR for the single source is a conventional
one: MONOGEM \cite{8}.

\section{Conclusions.}
That the suggested PSNR mechanism is very unlikely to work for even a single, rare 'object' is provided by the following facts.
\begin{itemize}
\item[(i)] The losses of energy by the pulsar-accelerated particles in collision with the very early SNR ejecta is expected to be very great. In our calculations we assumed that the ejecta was clumpy enough to allow such passage but this appears highly unlikely.
\item[(ii)] For both models considered here the 'source activity time' is far too long for any known type of pulsar to be responsible; CR particles emitted after the SNR illumination has ceased will dilute the spectral sharpness to an unacceptable degree. 
This is intrinsic in Model 1 and, for Model 2, can be seen directly in Figure \ref{fig7};for a typical cooling time of
100 d the pulsar must be active for less than a year.
\item[(iii)]Even if an ultra-rare pulsar of the required characteristics was proposed, the probability of its being close enough (and its contemporary presence undetected) is remote in the extreme.
\item[(iv)] The mass composition of the ambient CR in general and the Simgle Source in particular are quite different from what would be expected from a pulsar. 
%(unless it is the ambient interstellar medium that is subject to the 
%pulsar acceleration).
\end{itemize}

\section{Acknowledgements}
       The authors are grateful to Professors A. D. Erlykin and W. Bednarek
for helpful comments.

%\References

\newpage
%\section{Captions to the Figures.}

\end{document}